# Study and Capacity Evaluation of SISO, MISO and MIMO RF Wireless Communication Systems

Kritika Sengar[1], Nishu Rani[1], Ankita Singhal[1], Dolly Sharma[2], Seema Verma[1], Tanya Singh[2]

[1] Banasthali University, Newai, India

[2] Amity Institute of Information and Technology University, Noida, India

*Abstract* The wireless communication systems has gone from different generations from SISO systems to MIMO systems. Bandwidth is one important constraint in wireless communication. In wireless communication, high data transmission rates are essential for the services like tripple play i.e. data, voice and video. At user end the capacity determines the quality of the communication systems. This paper aims to compare the different RF wireless communication systems like SISO, MISO, SIMO and MIMO systems on the capacity basis and explaining the concept as today, the wireless communication has evolved from 2G, 3G to 4G and the companies are fighting to create networks with more and more capacity so that data rates can be increased and customers can be benefitted more. The ultimate goal of wireless communication systems is to create a global personal and multimedia communication without any capacity issues.

*Key Terms* MIMO, SISO, SIMO, MISO Capacity

## I. Introduction

In 1897, Guglielmo Marconi demonstrated first time the radio's ability to provide continuous contact with the sailing ships. Since then there had been tremendous development in wireless communication. Hundreds and thousands of engineers has worked and developed countless theories and researches on wireless communication [1]. During the last decade, the wireless communication industry has grown at exponential pace and people are taking more and better advantages of the technologies available from voice calling to video calling, from positioning to satellite television. The pace will continue in the upcoming future as well. For the users, the quality of the wireless communication can be defined by the availability and the data rates or capacity. Mobile communication starting from 2G, 3G and now 4G with the data rates varying from 12kbps in 2g to 2Mbps in 3g and 100Mbps in 4g  [2].

There has been significant in the data rates and the spectral efficiency of the radio wireless communication. The increase in data rates comes with the increase in the capacity of the systems.  In the very general, the mobile or wireless communication systems transmit information bits information in the radio space to the receiver. We are here going to discuss and simulate the capacity of systems like SISO systems, SIMO and MISO systems and MIMO systems [3].

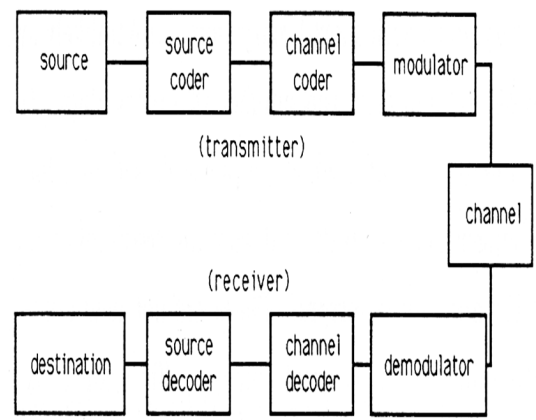

Figure 1 Wireless communication System

## II. SISO SYSTEMS

SISO Systems or the single input, single output communication systems is the simplest form of the communication system out of all four in which there is single transmitting antenna at the source and a single receiving antenna at the destination[3]. SISO systems are used in multiple systems like Bluetooth, Wi-Fi, radio broadcasting, TV etc.

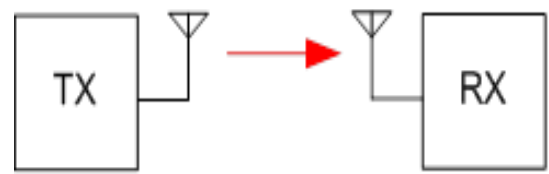

Figure 2 SISO Communication System

The capacity of such systems is given by Shannon capacity theorem giving the mathematical form as

$$C = B * \log_2(1 + S/R) \text{ bit/s} \ldots\ldots(1)$$

Where c = capacity, b= bandwidth of the systems, s/r is the signal to noise ratio.

SISO are advantageous in terms of the simplicity. It does not require processing in terms of diversity schemes. The throughput of the system depends upon the channel bandwidth and signal to noise ratio. In some conditions, these systems are exposed to the issues like multipath effects.

When an electromagnetic wave interacts with hills, buildings and other obstacles, waveform get scatter and takes many paths to reach the destination. Such issues are known as multipath. This causes several issues like fading, losses and attenuation also the reduction in data speed, packet loss and errors are increased.

### III. SIMO Systems

SIMO or the Single input and multiple output form of wireless communication scheme in which there are multiple antennas are present at the receiver and there is single transmitting antenna at the source. In order to optimize the data scheme, various receive diversity schemes are employed at the receiver like selection diversity, maximum gain combining and equal gain combining schemes. SIMO systems were used for short waves listening and receiving stations to counter the effects of ionosphere fading. The SIMO systems are acceptable in many applications but where the receiving system is located in the mobile device like mobile phone, the performance me be limited by size, cost and battery.

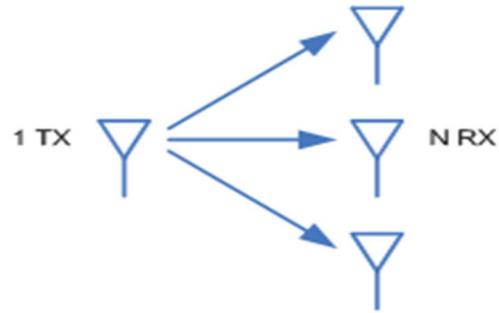

Figure 3 SIMO Communication Systems

### IV. MISO System

MISO or the multiple input and single output is a scheme of RF wireless communication system in which there are multiple transmitting antennas at the source and single receiving antenna at the system like SIMO but at the destination, receiver has a single antenna[4].

When we use two or more antenna at the receiving end or at destination, the effects of multipath wave propagation, delay, packet loss etc can be reduced. This scheme has various applications like in Digital television, W-lans. MISO systems are advantageous because the redundancy and coding has been shifted from receiving end towards the transmitting end and hence say in examples of mobile phones, less power and processing is required at the user end or the receiver end[5].

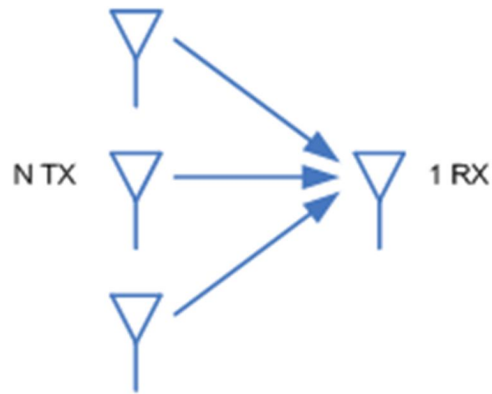

Figure 4 MISO Communication System

The capacity of MISO and SIMO systems can be expressed as

$$C = B * \log_2(1 + nS/R) \text{ bit/s} \ldots\ldots\ldots(2)$$

Where

n = number of transmit antenna in case of MISO systems and no. of receive antenna in case of SIMO systems.

C= Capacity of the system, B= Bandwidth of the system and S/R= Signal to noise ratio

## V. MIMO

MIMO systems or the multiple input and multiple output systems are the one with multiple antennas at transmitting end and multiple antennas at receiving end as well [6].

Between a transmitter and receiver, signal can go through many paths and if we move the antenna with a small distance, the path used by the signal will change.

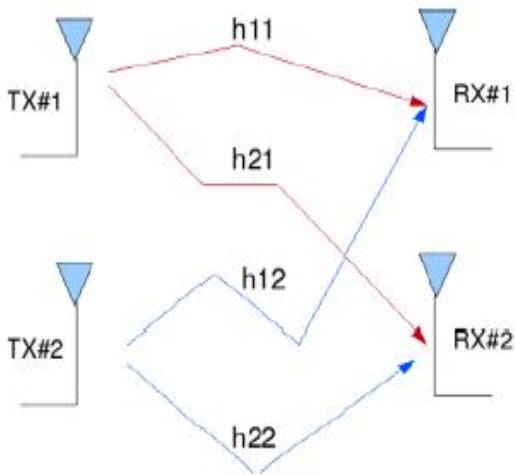

Figure 5 2x2 MIMO system

With the use of MIMO technology, the different paths available can be used for an advantage.[7]
By using MIMO, these additional paths can be used to advantage. They can be used to provide additional robustness to the radio link by improving the signal to noise ratio, or by increasing the link data capacity[8].
The capacity of the MIMO systems are given by the relation

$$C = B * \log_2(1 + nT.nR.S/R) \text{ bit/s} \ldots\ldots..(3)$$

Where,
nT= transmitter antenna,
nR= receiver antenna

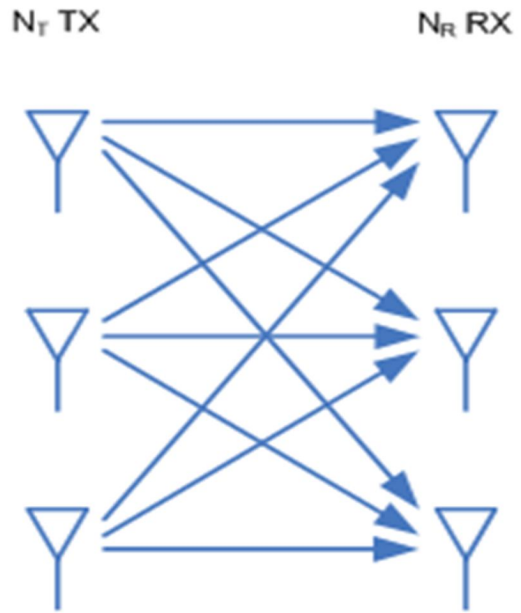

Figure 6 nxn MIMO System

But, with the signal is coded using techniques called space-time coding

$$C = \min(nT, nR) * B * \log_2(1 + S/R) \text{ bit/s} \ldots\ldots..(4)$$

Where,
Min (nT, nR) = minimum of nT and nR

MIMO systems often employs Spatial Multiplexing which enable signal to be transmitted across different spatial domains. MIMO is a hot topic in today wireless communications since all wireless technologies like PAN, LAN, MAN, and WAN trying to add it to increase data rate multiple times to satisfy their bandwidth-hungry broadband users[9][10][11].

## VI. Simulation

The equations defined above for capacity of the systems i.e. equation 1,2 and 3 are implemented in Matlab 7.10.0 for the simlation. For the sake of simulations, the value of B is taken as 1.

The following graph is showing the capacity if different MISO or SIMO systems with 2x1 or 3x1 configuration. The results for 1x2 and 1x3 will be same respectively as well.

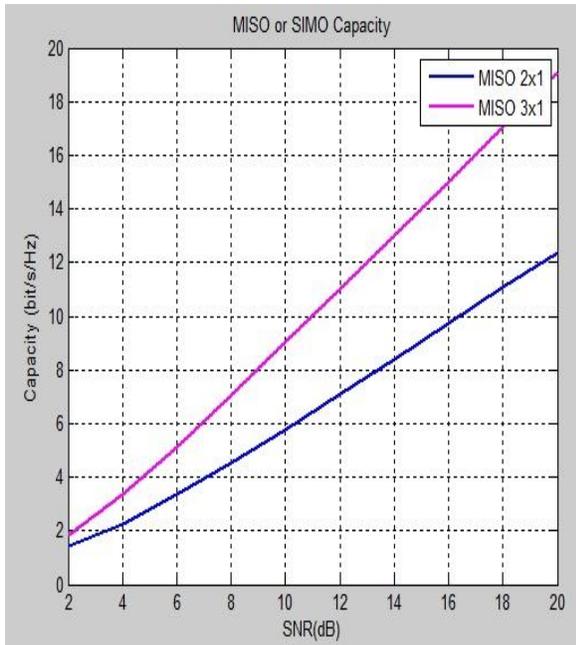

Figure 7 MISO System capacity for 2x1 ad 3x1

The capacity for SISO system and MIMO systems are shown and compared in the following graph

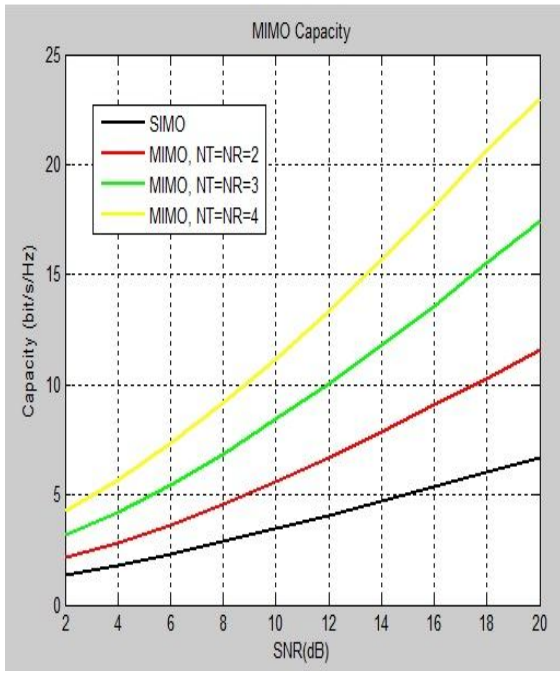

Figure 8 Capacity of SISO and MIMO systems

We can see that as the number of transmit and receive antenna increases, capacity is increasing. The black line is showing the capacity of SISO system and red, green and yellow lines show the system capacity for 2x2, 3x3 and 4x4 MIMO systems respectively.

In the following figure, all the three schemes are compared in different configurations i.e. SISO systems as 1x1, MISO or SIMO systems as 2x1 and 3x1 or 1x2 and 1x3 respectively and MIMO systems with 2x2 and 3x3 configurations.

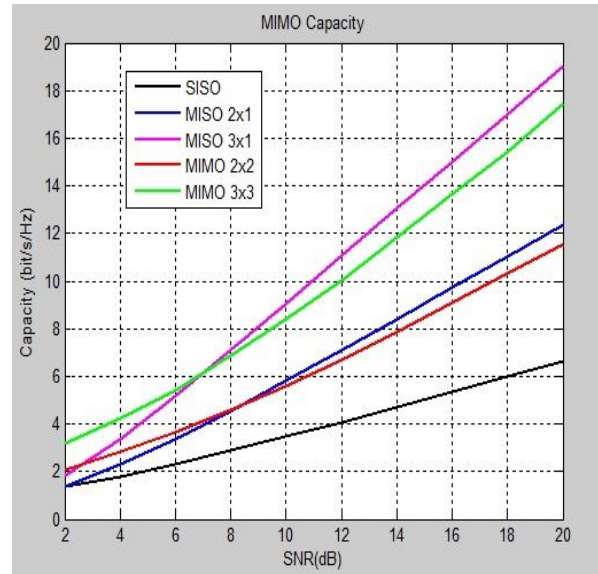

Figure 9 SISO vs MISO vs MIMO System

The capacity of MIMO systems is better than siso and other systems. MISO or SIMO systems show higher capacity but at the cost of the high SNR which is undesirable in wireless communication system.

Conclusion

The MIMO system shows the maximum capacity theoretically which has been proved by simulations as well. Also as the number of transmit and receive antenna increase in MIMO systems there capacity increases as the green line is showing in the figure 9. Also the different systems SISO, SIMO, MISO and MIMO are studied thoroughly to provide a base for the upcoming researchers to pursue their research in this area.